**Rewritable Artificial Magnetic Charge Ice**

Yong-Lei Wang[1,2,*], Zhi-Li Xiao[1,3,*], Alexey Snezhko[1], Jing Xu[1,3], Leonidas E. Ocola[4], Ralu Divan[4], John E. Pearson[1], George W. Crabtree[1,5] and Wai-Kwong Kwok[1]

[1] Materials Science Division, Argonne National Laboratory, Argonne, Illinois 60439, USA

[2] Department of Physics, University of Notre Dame, Notre Dame, Indiana 46556, USA

[3] Department of Physics, Northern Illinois University, DeKalb, Illinois 60115, USA

[4] Center for Nanoscale Materials, Argonne National Laboratory, Argonne, Illinois 60439, USA,

[5] Departments of Physics, Electrical and Mechanical Engineering, University of Illinois at Chicago, Chicago, Illinois 60607, USA

* Correspondence to: ylwang@anl.gov and xiao@anl.gov

Abstract:     **Artificial ices enable the study of geometrical frustration by design and through direct observation. It has, however, proven difficult to achieve tailored long-range ordering of their diverse configurations, limiting both fundamental and applied research directions. Here, we design an artificial spin structure that produces a magnetic charge ice with tunable long-range ordering of eight different configurations. We also develop a technique to precisely manipulate the local magnetic charge states and demonstrate write-read-erase multi-functionality at room temperature. This globally reconfigurable and locally writable magnetic charge ice could provide a setting for designing magnetic monopole defects, tailoring magnonics and controlling the properties of other two-dimensional materials.**

1      Artificial ices are structures in which the constituents obey analogues of the 'two-in two-out' Pauling's ice rule that determines the proton positional ordering in water ice; they provide a material-by-design approach to physical properties and functionalities (*1-27*). Artificial ice can be created from ferromagnetic islands and connected wires (*3,5*) as well as from topological components such as superconducting vortices (*24-26*) and from non-magnetic colloidal particles (*27*). Among them, artificial spin ice is the most investigated system (*1-21*) that was first demonstrated in a square lattice of elongated interacting ferromagnetic nano-islands (*1*). In this case, the ice rule corresponds to two spins pointing inward and two pointing outward at the vertex of a square lattice. Extensive experiments have been conducted using various thermal (*6-8*) and magnetization approaches (*9-11*) to obtain ordered states of the spin ice. Long-range ordering was realized in the diagonally polarized magnetic state through the magnetization method (*11,14*). For the nominal ground state, sizeable domains and crystallites were obtained in as-grown samples (*11*), and larger domains were obtained in samples heated above their Curie temperatures (~500 °C for permalloy island arrays) (*6*). At room temperature, long-range ordering of the ground state has only been achieved in ultrathin (3 nm thick) samples via thermal relaxation (*8*). Other spin/charge configurations have only been observed locally and at crystallite boundaries (*5,6*). The difficulty in creating tailored multiple ordered states limits the experimental investigation of the spin and magnetic charge dynamics that can emerge from/between the ordered states (*6, 10-13*), especially for a thermally stable (athermal) sample at room temperature. This also hinders the potential applications of artificial ice for data storage, memory and logic devices (*3,4*), or as a medium for reconfigurable magnonics (*28*).

2      Here, we design an artificial spin structure in which we can conveniently obtain multiple long-range orderings of a magnetic charge ice lattice at room temperature.

3      In a typical square spin ice (Fig. 1A), the single domain magnetic islands are considered as macro-Ising spins (*11*). Each 'macro-spin' can be replaced with a dumbbell of magnetic charges, one positive and one negative (*3,4,29*) (Fig. 1B). If we break the connections between the pairs of magnetic charges with opposite signs (Fig. 1C), we can design a different pattern of connections (Fig. 1D). Figure 1E shows the resulting artificial spin structure explicitly. The structure consists of a square lattice of plaquettes (labeled M and N) containing ferromagnetic nano-islands with three orientations (horizontal, vertical and diagonal). The magnetic charge distribution is indicated by the calculated stray field distribution (Fig. 1F), which leads to a magnetic charge ice with the charge ice rule of 'two-negative two-positive' charges within each square plaquette.

4      The magnetic charge distribution in Fig. 1F is exactly the same as that of the ground state of a square spin ice structure (see Fig. S1). However, in contrast to the square spin ice, which has spins on four sides of the square plaquette and all oriented towards the plaquette center/vertex (Fig. 1A), the structure in Fig. 1E places one spin within the plaquette while removing two from the sides, breaking the four-fold symmetry of a square lattice; moreover, the three spins associated with each plaquette do not meet at a vertex. Our design contains two types of plaquettes, one rotated by $180^o$ from the other, denoted as M and N in Fig. 1E. Each plaquette (M or N) consists of three islands, each with two degrees of freedom coming from spin, resulting in a total multiplicity of eight configurations of magnetic charges. See Figs. S1 and S2 for detailed comparisons of the spin/charge configurations between the standard square spin ice and our design. The predicted eight ordered spin/charge configurations (Fig. 1G) are separated into three groups based on their energies (see the calculated energies in Fig. S3): a two-fold degenerate Type-I ground state ($I_1$ and $I_2$), a two-fold degenerate excited Type-II state ($II_1$ and $II_2$), and a four-fold degenerate excited Type-III state (III$_1$, III$_2$, III$_3$ and III$_4$) (Fig. 1G).

5      Because of the large energy barrier for the spins, the spin system in Fig.1E is athermal at room temperature. An external applied magnetic field can be used to overcome the energy barrier. For a given island, the minimal magnetic field required to flip its spin moment varies with the angle between the applied field and the island (Fig. S4). This enables separate control of the spin moments of each differently oriented island. In a square spin ice, there are two orientations of the islands and the spins can only be aligned in diagonal directions by an applied magnetic field, enabling only the Type-II phases with long-range ordering (*11,14*). Our design (Fig.1E) provides an additional freedom for aligning the spins: there are three sets of islands orientated in the horizontal, vertical and diagonal directions. More importantly, for each of the predicted ordered states shown in Fig. 1G, the spin moments of each of the three oriented islands (horizontal, vertical or diagonal) are all magnetized in the same direction. This enables the creation of long-range ordering for all the predicted charge configurations by tuning the in-plane external magnetic field angle and amplitude. We calculated the field angle dependence of the moment flipping curves for the three sets of islands (Fig. S4C), and designed an effective magnetization protocol to realize the various ordered charge states (*30*).

6      To experimentally demonstrate the magnetic charge ordering, we fabricated arrays of permalloy (Ni$_{0.8}$Fe$_{0.2}$) nanoislands (300 nm long, 80 nm wide, and 25 nm thick) onto a Si/SiO$_2$ substrate (*30*) according to the design in Fig.1E. A scanning electron microscopy (SEM) image of the sample is presented in Fig.2A. To control and visualize the charge configurations, we used a customized magnetic force microscope (MFM) equipped with a 2D vector magnet. Figure 3A shows a schematic drawing of our experimental setup. The 2D electromagnetic solenoid magnet provides in-plane magnetic fields in any desired orientation, enabling us to accurately tune the field angle and amplitude.

7       The as-grown (Fig. S5) and demagnetized (Fig. S6) samples show mixed charge states of all the eight charge configurations at both M and N plaquettes. Statistical analysis for the demagnetized samples shows that the charge neutral configurations (Type-I) are strongly favored and the collective interaction can be enhanced by reducing the charge separation (Fig. S6). Using the designed magnetization protocol (Fig.S7), we successfully obtained all eight configurations of magnetic charge ordering, as shown by the MFM images in Fig. 2, B-I. Each of these ordered states possesses long-range ordering and can be reproduced over the entire patterned sample area (80 x 80μm$^2$). For Type-I and Type-II ordered states, there is no net charge in each plaquette and the magnetic charge follows the "two-positive two-negative" charge ice rule. In the Type-III ordered states, each plaquette with magnetic charges of "three-positive (negative) one-negative (positive)" has an effective magnetic charge of two with opposite signs on the M and N plaquettes. This distribution of magnetic charges in the square lattice of the Type-III state resembles the distribution of electrical charges in ionic compounds, such as $Mg^{2+}O^{2-}$. Thus, the Type-III states (Fig. 2, F-I) resemble an ionic crystal with magnetic charges (*31*) where we can associate $M^{2+}N^{2-}$ with the Type-III$_2$ and Type-III$_4$ state and $M^{2-}N^{2+}$ with the Type-III$_1$ and Type-III$_3$ state. In principle, all these magnetic charge distributions could also exist in a square spin ice structure (Fig. S1E and Fig. S2). In fact, our micromagnetic simulation result indicates that the square spin ice and our engineered spin structure not only produce the same magnetic charge arrangements but also have the same excitation energies for the Type-I, II and III configurations (Fig. S3). However, with the square spin ice arrangement, long range ordering of Type-I states has only been realized in thermally relaxed sample and that of the Type-III states has not yet been experimentally realized because of the spin arrangement of the islands.

8 The reconfigurable ordered magnetic charges can be used, for example, as templates to form other artificial ices such as superconducting vortex ices (*25,26*) by introducing ice-like pinning potentials for superconducting vortices. It can also be applied to couple with other electronic materials, such as two-dimensional electron gas (*32,33*) and graphene (*34*), by producing reconfigurable periodically distributed field potentials. For applications such as data storage, memory and logic devices (*3,35,36*), however, local control of the magnetic charge states is desired. Toward this end we developed a 2D magnetic field assisted MFM patterning technique, which allows us to conveniently manipulate the local charge configurations.

9 As illustrated in Fig. 3A, the magnetic tip of an MFM generates an in-plane component of stray magnetic fields near the tip. The interaction of the MFM tip's stray field, $\Delta H^m$, with a single ferromagnetic island can be tuned by adjusting the height of the MFM tip from the sample, which is 100 nm in this experiment. To locally switch the spin states of an island, we apply an in-plane magnetic field $H^{ap}$ slightly below the ferromagnetic island's spin moment flipping field $H^f$ (Fig. 3B). At this field value the spin states of the entire sample will not be altered since $H^{ap} < H^f$. When the MFM probe scans over an island, the total magnetic field on that island will change in the range $H^{ap} \pm \Delta H^m$. We adjust $H^{ap}$ and $\Delta H^m$ to satisfy the condition $H^{ap} < H^f < (H^{ap} + \Delta H^m)$ (light blue region in Fig. 3B). In this case, the spin of the underlying island flips when the MFM tip scans over it, providing a 'write' function. A subsequent applied magnetic field in the opposite direction with $-(H^{ap} + \Delta H^m) < -H^f < -(H^{ap} - \Delta H^m)$ (green region in Fig. 3B) will switch the spin back, implementing the 'erase' function (Fig. 3B). When the applied field is zero, the stray field provided by the MFM probe (yellow region in Fig. 3B) is too small to flip any islands, and this works as the 'read' mode. Because the value of $H^f$ depends on the angle between the applied field and the island, similar to the global control of the charge ordering, we can locally manipulate the charge states into any desired configurations.

10      We demonstrate the experimental realization of the 'write', 'read' and 'erase' functions in Fig. 3,C-F. We first prepare the entire sample in the Type-I ground state (Fig. 3C) by applying and zeroing an in-plane magnetic field of 90 mT along the diagonal direction. We then 'write' a square area of Type-III ordered state in the center (Fig. 3D) and subsequently 'erase' a smaller square area in the central area by switching the Type-III state back into the Type-I state (Fig. 3E). Finally, we 'write' a Type-II state into a small circular area inside the Type-III square (Fig. 3F). We can also write letters, as presented in Fig.3G where the word 'ICE' is scribed with Type-III order on a Type-I background. Such magnetic charge patterning can be easily realized by programming the 2D magnet to turn on/off and to switch the field directions during the MFM scanning, resembling the patterning process used in electron-beam lithography and in optical lithography using a laser pattern generator. We demonstrate more patterns of magnetic charge arrays in Fig. S8. These rewritable magnetic charge patterns could be transferred to other materials, for example, through magnetolithography (*37*).

11      In addition to the aforementioned applications on coupling with superconducting vortices and other electronic systems, this reconfigurable magnetic charge ice can provide a unique platform to explore phenomena such as the ground state of a frustrated lattice (*6*). Combined with other control parameters such as the thickness of the islands (*11*), temperature (*6*) or oscillating magnetic field (*10*), our platform provides a versatile system to study and tailor phase transitions and defect formation (Fig. S9). For example, the single spin control enabled by our method allows the creation of magnetic defects such as magnetic monopoles and Dirac strings (*16,17,20*) at any desired locations. It also provides a direct technique to program magnetic logic circuits (*35,36*). Our strategy to decouple the arrangement of spins and magnetic charges should stimulate further creation and exploration of exotic phases of magnetic charges, their phase transitions and also foster applications. We also note that Fig.1E is not the only spin arrangement for achieving the same magnetic charge distribution. In Fig. S10, we present several other possible designs of the spin/charge arrangements, including the Type-IV charge ordered state where all positive or negative charges are confined within a single plaquette. Furthermore, it is not necessary to keep the length of all the ferromagnetic islands the same, as recently reported in Shakti spin ices (*18,19*). Our strategy could also be applied to other artificial spin ices to produce artificial structures with controllable magnetic charge orders, which would provide reconfigurable platforms for magnonic investigations, such as programming spin-wave band structures and designing spin-wave transmission channels (*5,28*).

**Acknowledgments:** We thank W. J. Jiang, S. Zhang, W. Zhang and M. P. Smylie for critical comments. This work was supported by the U.S. Department of Energy, Office of Science, Basic Energy Sciences, Materials Sciences and Engineering Division. Z. L. X. and J. X. acknowledge NSF Grant No. DMR-1407175.  Use of the Center for Nanoscale Materials, an Office of Science user facility, was supported by the U. S. Department of Energy, Office of Science, Office of Basic Energy Sciences, under Contract No. DE-AC02-06CH11357.

**Fig. 1 Design of magnetic charge ices.** (A) Typical artificial square spin ice in which the length of the magnetic nano-island equals the separation of the ends of the nearest islands. Each island is an Ising spin (black arrows). Shown is the spin configuration of the lowest energy ground state. (B) Distribution of magnetic charges corresponding to the square spin ice shown in (A). The pairing of the positive (red) and negative (blue) charges is indicated by black dotted lines. (C) Distribution of magnetic charges with charge connections removed. (D) Redesign of the connections of paired charges with positive and negative polarities. (E) The design of the magnetic nano-structure based on (D). The arrows indicate the spin configuration of the ground state and are color coded for the three subsets of islands placed in three different orientations. Two types of plaquettes are marked with M and N. (F) Calculated magnetic stray field associated with the spin configuration of the structure in (A) for islands with dimensions 300 nm × 80 nm × 25 nm. The arrows indicate the local field directions. The color is encoded by the out-of-plane component of the field (red, out of the plane; blue, into the plane). The red and blue spots represent the positive and negative magnetic charges, respectively. (G) Calculated magnetic stray fields of a 2 × 2 square plaquette, highlighted by the red frame in (E), for the eight possible spin/charge ordered configurations, separated into three charge types. The dotted lines denote the orientation of the islands containing the pair of magnetic charges (negative charge: blue; positive charge: red). In order to compare the magnetic simulations with the experimental MFM images, the stray field is calculated at a plane 100 nm above the surface of the sample.

**Fig. 2 Realization of magnetic charge ices.** (A) Scanning electron microscopy (SEM) image of permalloy ($Ni_{80}Fe_{20}$) magnetic islands (300 nm long, 80 nm wide, and 25 nm thick). (B-I) Magnetic force microscopy (MFM) images of the various ordered states corresponding to all the configurations in Fig. 1G: (B, C) two-fold degenerate Type-I ground states: $I_1$ (B) and $I_2$(C); (D, E) two-fold degenerate excited Type-II states: $II_1$ (D) and $II_2$ (E); (F-I) four-fold degenerate excited Type-III states: $III_1$ (F), $III_2$ (G), $III_3$ (H) and $III_4$ (I). The lift height of MFM scanning is 100 nm.

**Fig. 3 Rewritable magnetic charge ices.** (A) Sketch of the experimental setup: a magnetic force microscope (MFM) equipped with a 2D vector magnet. The 2D solenoid magnet provides magnetic fields in any desired orientation in the sample plane. Vertically magnetized MFM probe generates stray magnetic field (green arrows) with in-plane components at the tip. (B) Magnetization loop of a single magnetic island with illustration of the 'write', 'erase' and 'read' functions. (C-G) MFM images of the patterned magnetic charge ice at the same area of the sample: (C) initial state is a Type-$I_1$ state; (D) a square area of Type-$III_3$ state was written in the center of (C); (E) a smaller square region of Type-$III_3$ order is 'erased' back to Type-$I_1$ state from (D); (F) a round region of Type-$II_2$ order is 'written' onto the freshly 'erased' area from (E); (G) 'ICE' letters of Type-$III_4$ states were scribed on a Type-I background state.

**Supplementary Materials:**
Materials and Methods
Figures S1-S10
Reference *(38)*

**Fig. 1:**

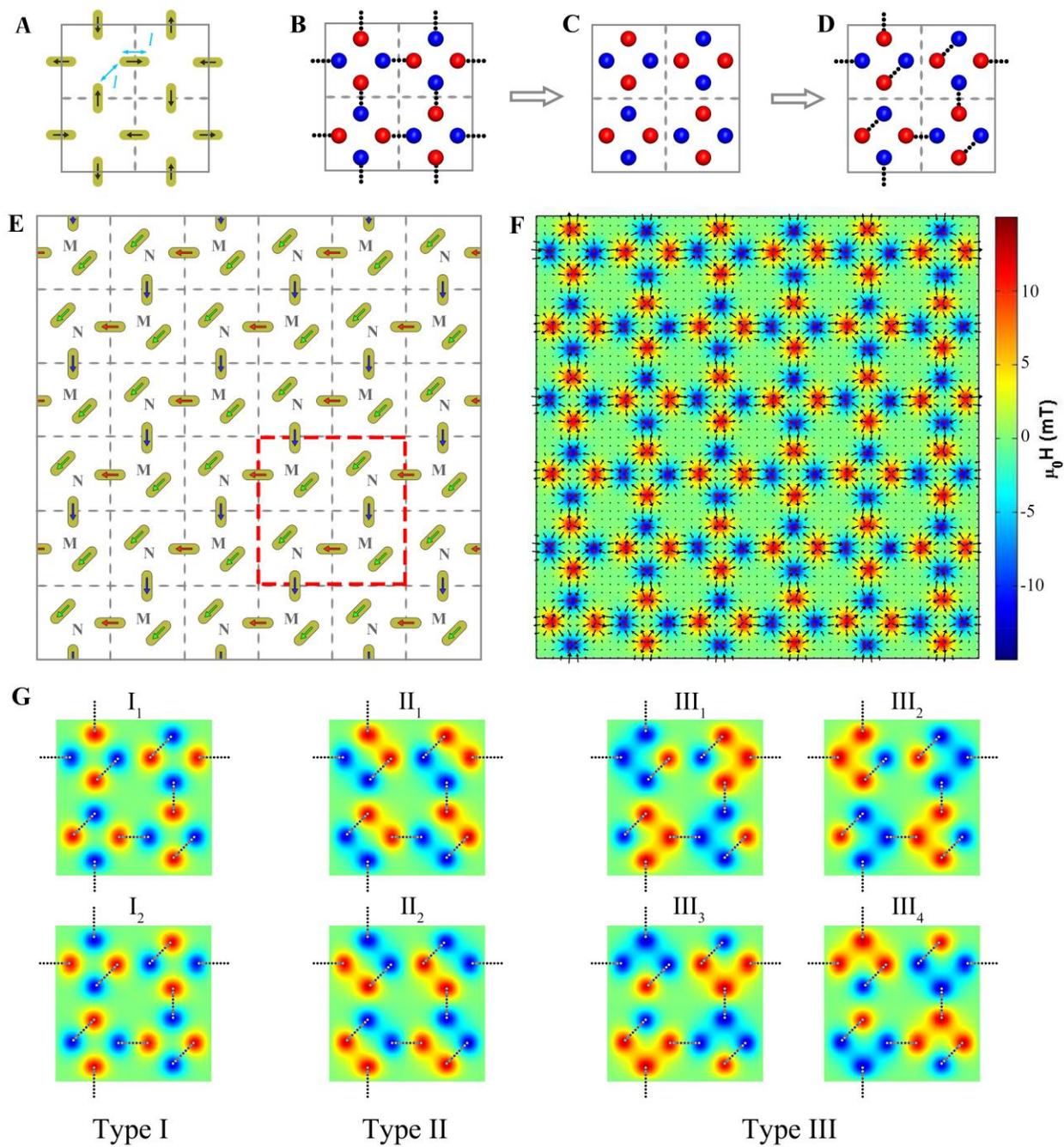

Fig. 2:

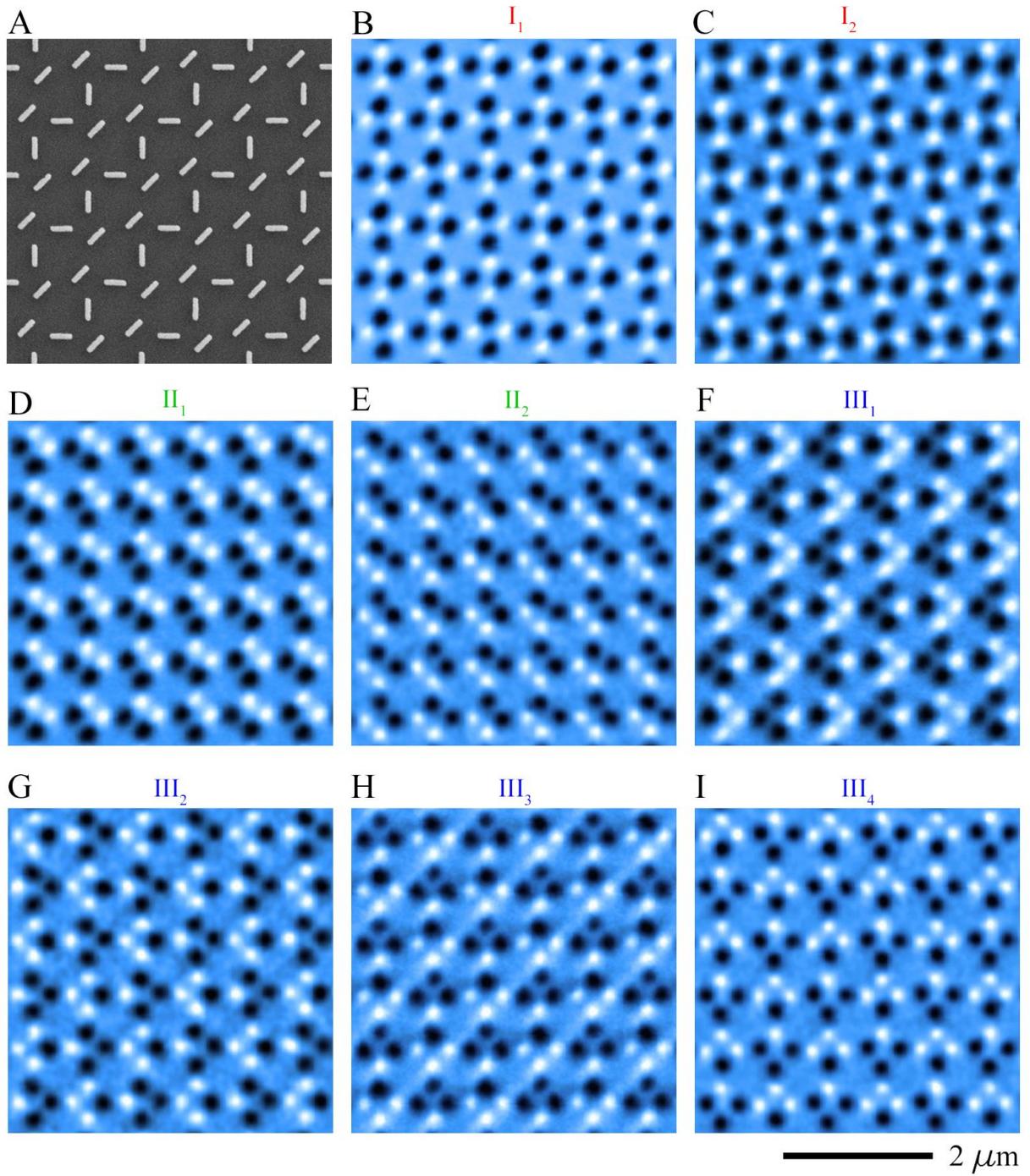

**Fig. 3:**

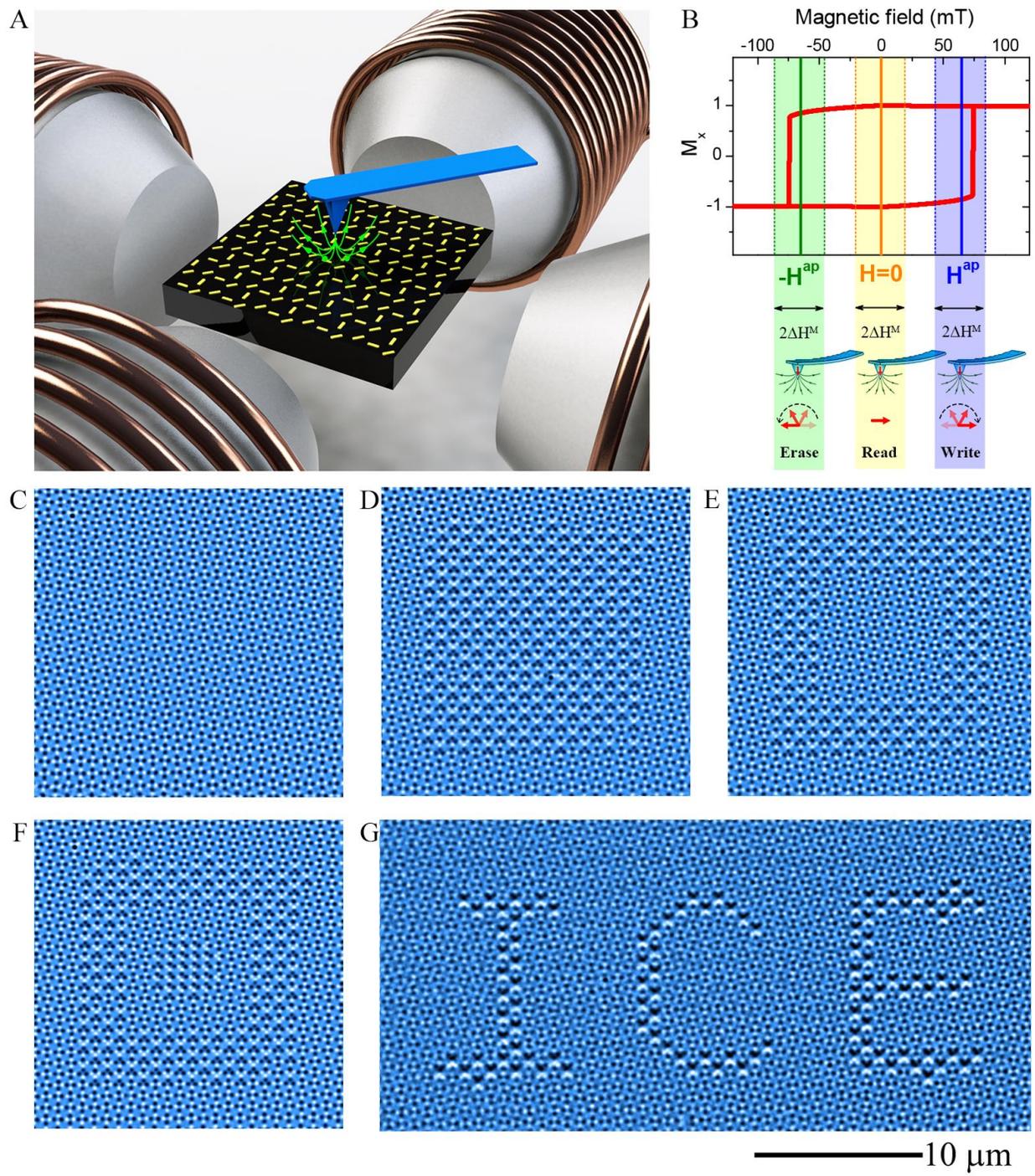

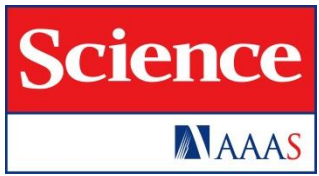

# Supplementary Materials for

## Rewritable Artificial Magnetic Charge Ice

Yong-Lei Wang*, Zhi-Li Xiao*, Alexey Snezhko, Jing Xu, Leonidas E. Ocola,
Ralu Divan, John E. Pearson, George W. Crabtree and Wai-Kwong Kwok

*correspondence to: ylwang@anl.gov and xiao@anl.gov

**This PDF file includes:**

    Materials and Methods
    Figs. S1 to S10

**Other Supplementary Materials for this manuscript includes the following:**

**Materials and Methods**

Sample fabrication.

We fabricated the magnetic nano-island arrays on Si/SiO$_2$ substrates using films of permalloy (Ni$_{0.80}$Fe$_{0.20}$). We employed a lift-off technique using double layers of polymethyl methacrylates (PMMAs). First, a 495 PMMA resist layer with thickness about 80 nm was spin-coated onto the Si substrate at 2000 rpm for 45 seconds. After baking at 180 $^{\circ}$C for 3 minutes and cooling down to room temperature, a 950 PMMA resist layer with thickness of about 100 nm was spin-coated at 2000 rpm for 45 seconds, followed by baking at 180 $^{\circ}$C for 3 minutes. The sample was exposed to an electron-beam with a dose of 350 μC/cm$^2$ and energy of 30 keV. Subsequently, it was placed in a PMMA developer (MIBK:IPA ratio = 1:3) for 50 seconds and in IPA for 30 seconds. We used an electron beam evaporator to deposit a 25 nm thick permalloy film at a rate of 0.3 angstrom per second. The lift-off process was carried out in a '1165' remover at 70 $^{\circ}$C for 2 hours, followed by 2 minutes of ultrasonic shaking. After cleaning in IPA, the process yielded nanometer-scale permalloy islands with dimensions of 300 nm long, 80 nm wide and 25 nm thick on the Si/SiO$_2$ substrate. The final array size is 80 μm × 80 μm with a total of 24,200 islands.

Micromagnetic simulation.

Simulations of magnetization loops, spin flipping curves, stray field distributions and energies of various charge ordering were carried out using the Object Oriented Micromagnetic Framework (OOMMF) Code (*38*). The permalloy parameters used in our simulations were: exchange constant A = 1.3 × 10$^{-11}$ J/m, saturation magnetization M$_s$ = 8.6 x 10$^5$ A/m, magnetocrystalline anisotropy constant K = 0. The cell size is 5 nm × 5 nm × 5 nm. The excitation energy for infinite sized sample is calculated for a 2 × 2 lattice of plaquettes under 2D periodic boundary conditions for both the designed spin structure and the corresponding square spin ice structure. The stray field distribution is also calculated under 2D periodic boundary conditions.

MFM imaging.

The experiments were conducted in a custom designed and built MFM system placed in a 2D electromagnet. The magnetic charge imaging and manipulation were implemented using a commercial MFM probe (NANOSENSORS™ PPP-MFMR), where the tip was vertically magnetized. To avoid inadvertent flipping of the magnetic island's moment, we used a constant height scanning method for MFM imaging with the scanning plane set at 100 nm above the sample. All MFM images were obtained in zero applied field.

Manipulation of the magnetic charge ordering with magnetic field.

Figure S4A shows a typical calculated magnetization loop of a ferromagnetic island. When the applied magnetic field is larger than the flipping field, H$^f$, the island switches its spin direction. Since the value of H$^f$ depends on the applied field direction, as shown in Fig. S4, it is possible to selectively control the switching of the spin moments of the islands placed in different orientations by tuning the angle and amplitude of the applied field. In Fig. S4C we plot the angle dependence of the flipping fields for the three subsets of islands/spins in polar coordinate.

To realize a specific configuration of charge order, we refer to the flipping field curves shown in Fig. S4. When the applied magnetic field crosses the flipping field curve of the islands in a specific orientation, the moment of the corresponding islands will flip if it initially points in the opposite direction with respect to the applied field. We present procedures to obtain all eight charge configurations in Fig. S7. The magnetic moments of the vertical and horizontal islands can be simply flipped by applying an in-plane field to $H_2/H_{4,8}$ and $H_3/H_7$, respectively, as shown in Fig. S7A. Flipping the diagonally oriented islands requires a two-step procedure because there is no direct pathway to apply a field that only crosses the flipping curve of the diagonally oriented islands (the green curve in Fig. S4C and Fig. S7A). The method is to flip both the diagonal islands and one set of the horizontal or vertical islands first, and then flip the latter back. For example, in Fig. S7A, applying a field to $H_5$ will flip the spins of both the diagonal islands and the horizontal islands, and then applying a field to $H_7$ will flip the moment of the horizontal islands back to its original direction, with the final result of only flipping the moment of the diagonal islands.

Sample demagnetization

In-plane magnetic field rotating at 20 rpm starts at 1000 Oe (above the coercive field of the islands) and gradually reduces to zero in steps of 0.1 Oe at a rate of 3.33 steps per second (300 ms/step). A small AC field of 5 Oe was superimposed at a higher rotation rate of 100 rpm to assist the fluctuation of the macrospins.

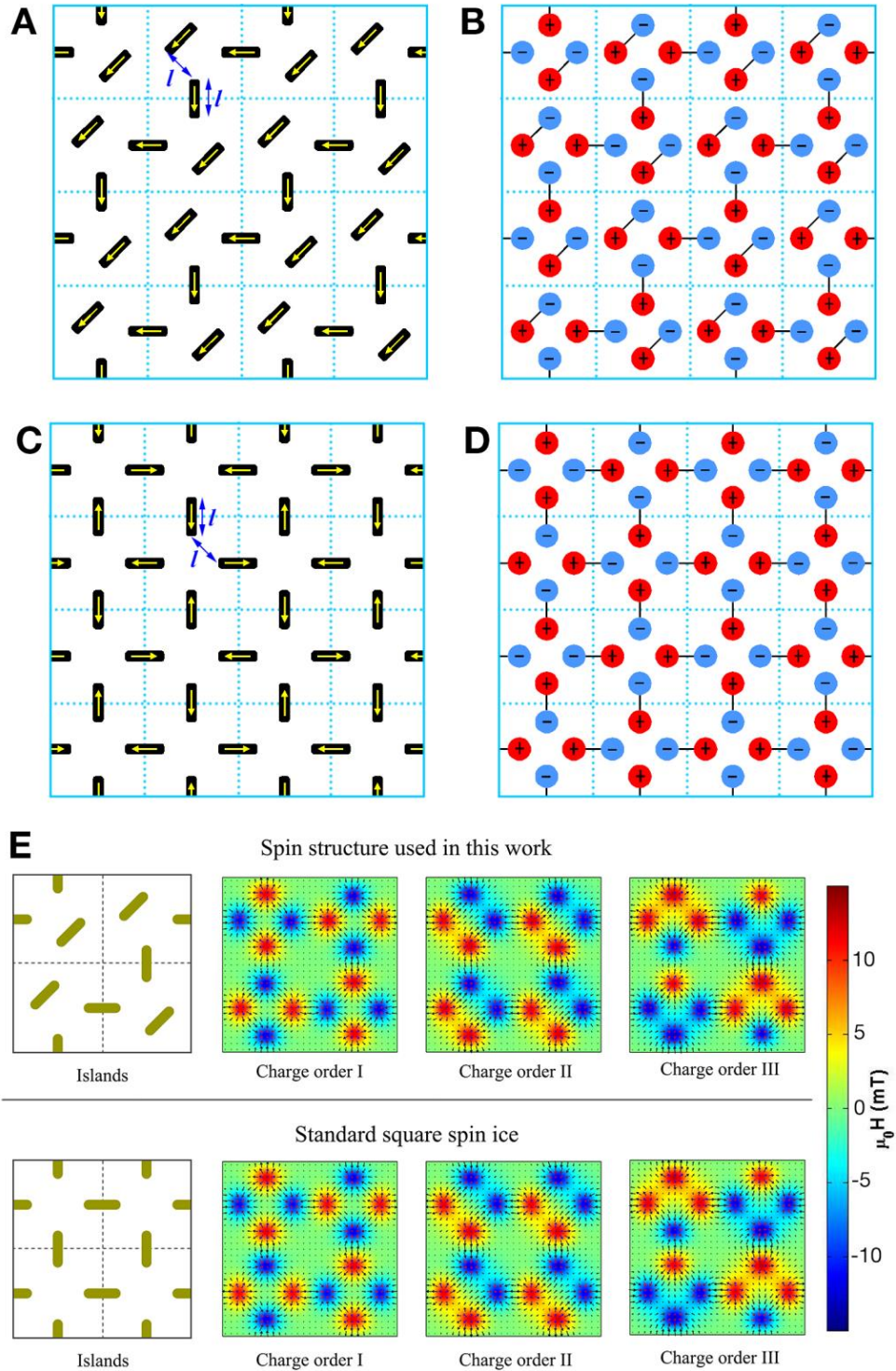

**Fig.S1. Distributions of the spins and magnetic charges in the designed artificial spin structure and an artificial square spin ice structure**. (A) and (C) illustrate the nanomagnet arrangements of the designed spin structure and a square spin ice structure, respectively. The nearest neighbor distance equals the length $l$ of the nanomagnets. The spins, indicated by arrows, in (A) and (C) correspond to the configurations in the ground

states. (B) and (D) show the magnetic charge ice distributions associated with the spin arrangements in (A) and (C), respectively. The red and blue dots correspond to positive and negative magnetic charges. The black lines in (B) and (D), connecting pairs of positive and negative magnetic charges, mirror the islands in (A) and (C), respectively. (E) Calculated magnetic stray fields of a $2 \times 2$ square plaquette for both the designed spin structure (top panels) and the square spin ice structure (bottom panels) with three corresponding charge states (Type-I, Type-II and Type-III) in both structures. The arrows indicate the local field directions and are color coded with the out-of-plane component of the field (red, out of the plane; blue, into the plane). The red and blue spots represent the positive and negative magnetic charges, respectively. The magnetic charge distributions of the two structures are equivalent while their spin arrangements differ.

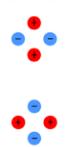

**Fig.S2. Comparison of the local charge and spin configurations of the artificial square spin ice and the spin structure presented in this work.** The percentages indicate the expected fraction of each type in a non-interacting island array.

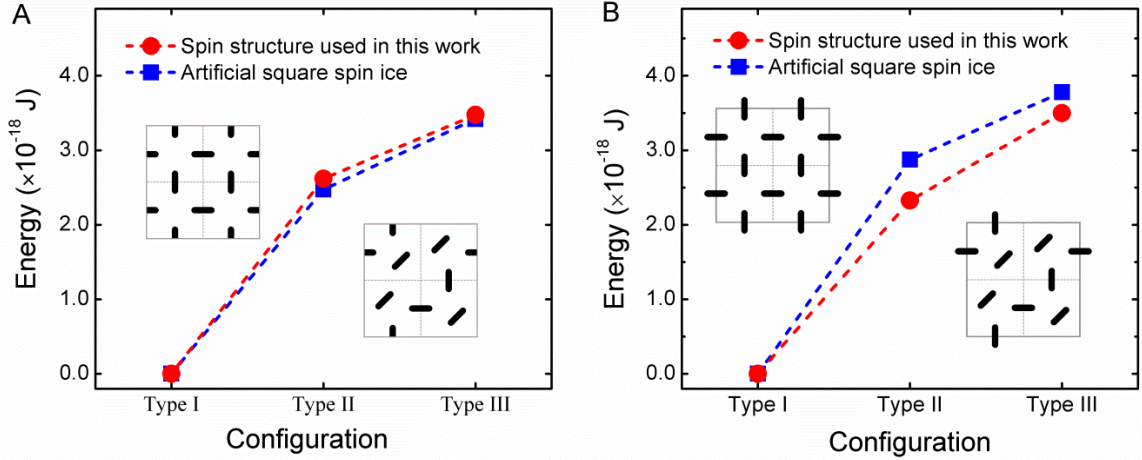

**Fig.S3. Calculated excitation energies of an artificial square spin ice and the spin structure presented in this work.** (A) for infinite-sized samples and (B) for small samples with 2 by 2 lattices. The dimensions of the islands used in the calculation are 300 nm × 80 nm × 25 nm with round ends. Calculations were conducted using the islands shown in the insets in (A) and (B). The calculations for infinite-sized samples in (A) were conducted under 2D periodic boundary condition.

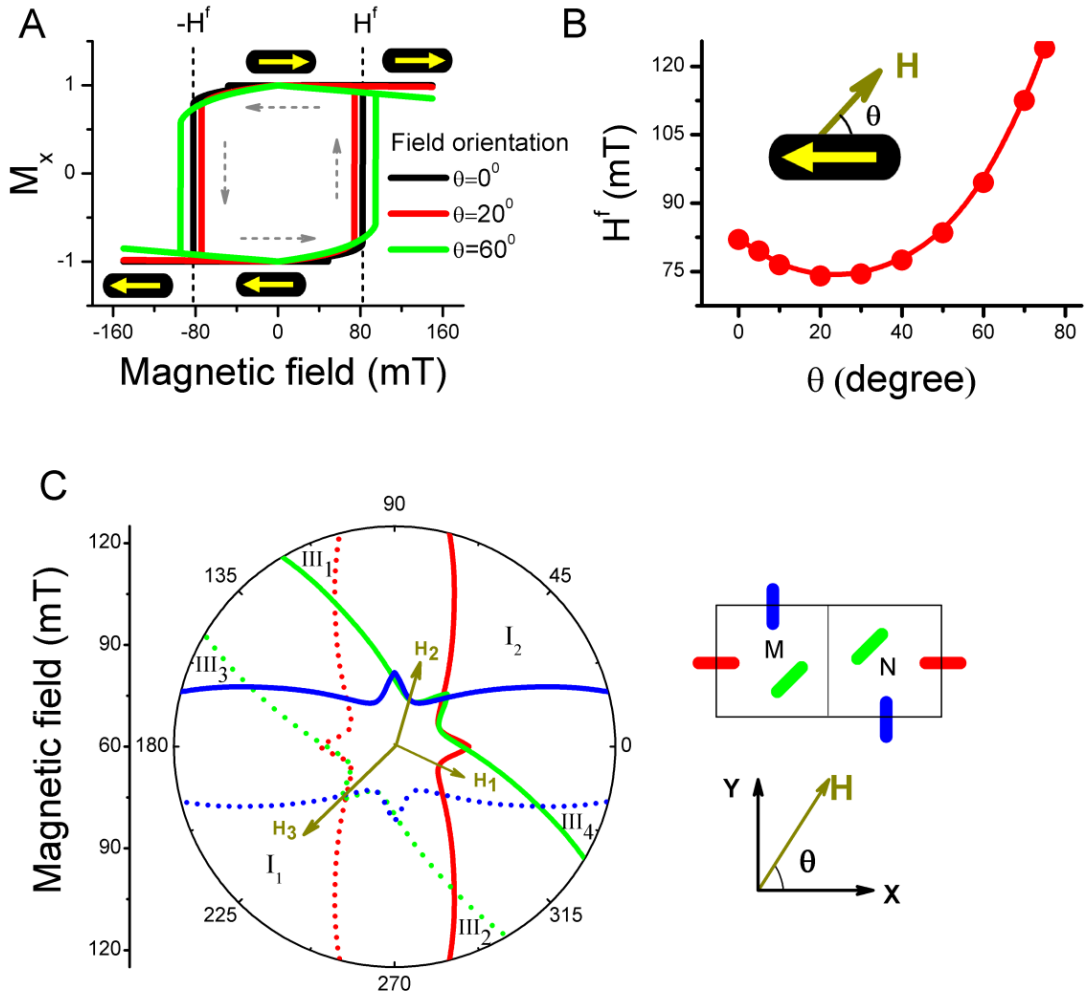

**Fig.S4. Angle dependence of the moment flipping fields of the magnetic islands.** (A) Calculated magnetization loop of a Py island (300 nm long, 80 nm wide and 25 nm thick) with fields along $0^o$, $20^o$ and $60^o$ with respect to the long axis of the island. $H^f$ is the minimum flipping field at $\theta = 0^o$, above which the spin reverses its direction. The gray dashed arrows indicate the field sweep directions. The spin directions of the islands at various applied fields are shown by the yellow arrows. (B) Angle dependence of the flipping/switching field $H^f$ curve. The inset shows the applied field direction with respect to the island. For a single 300 nm × 80 nm × 25 nm structure, the most effective field direction to flip the spin moment is approximately $22^o$, and can change with the dimensions of the nano-islands. (C) Polar plots of the angle dependences of the flipping magnetic fields of the three subsets of islands orientated horizontally (red curves), vertically (blue curves) and diagonally (green curves). Dotted curves and solid curves represent switching field values of the three oriented islands. The six flipping curves separate the field coordinate space into seventeen sections. For field vectors within the center star-region, the initial spin/charge configurations will not change. When the applied field vector crosses the flipping curves, the corresponding oriented islands flip their spin moment. For example, for field vector $H_1$, which crosses the red solid line

flipping curve (for horizontal islands), the spins of the horizontal islands point to the right. Depending on the initial spin/charge states, the charge configuration can be Type-$I_2$, $II_2$, $III_2$ or $III_4$. For the field vector $H_2$, which crosses two flipping curves (for vertical and diagonal islands), the spins of the vertical and diagonal islands point up and up-right, respectively. The charge configuration can be either Type-$I_2$ or $III_1$, depending on the initial states of the spins in the horizontal islands. For field vector $H_3$, which crosses three flipping curves, there is only one charge configuration, Type-$I_1$, and in the figure we mark all such field regions with their corresponding charge configurations, which indicates that all the Type-I and Type-III states can be easily obtained through a single magnetization step. The realization of Type-II states depends on the initial charge states (see Fig. S7 for details on obtaining all the various charge ordered configurations).

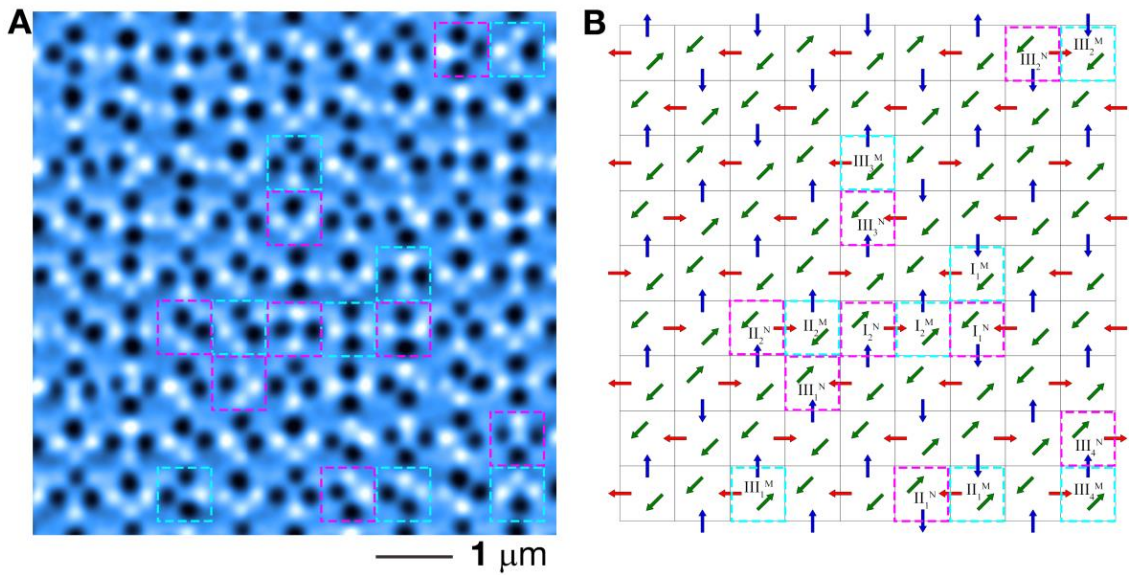

**Fig.S5. Magnetic charge/spin configurations of an as-grown sample.** (A) Magnetic force microscopy image of an as-grown sample. (B) The magnetic moments (arrows) extracted from (A). Eight charge configurations are identified by the dashed square frames of M (sky blue) and N (purple) type plaquettes in both (A) and (B). The corresponding charge-order configuration labels are labeled in (B).

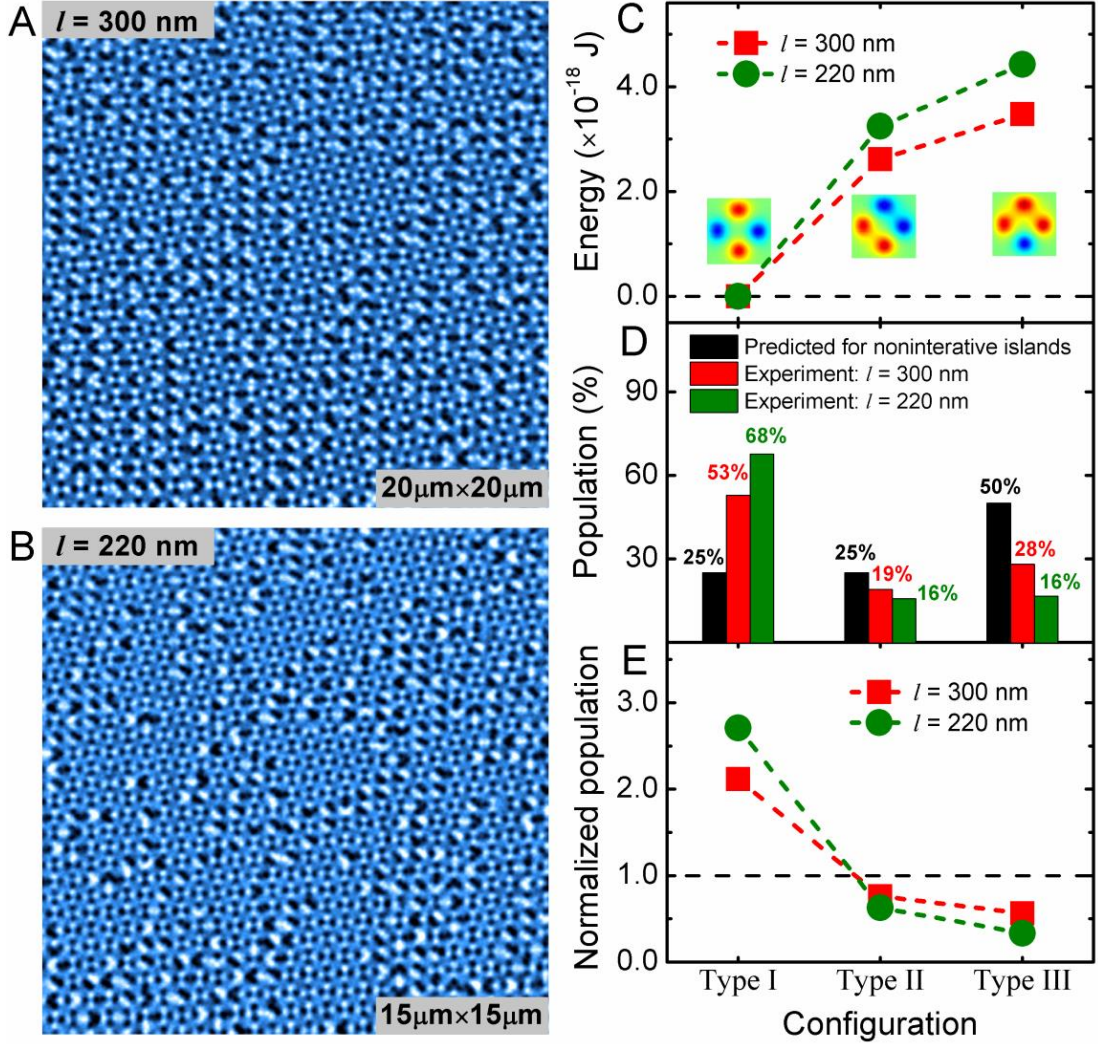

**Fig.S6. Collective behavior of magnetic charge ice.** (A, B) MFM images of demagnetized states with charge spacing of $l = 300$ nm (A) and $l = 220$ nm (B) (see Fig.S1A for the definition of $l$). The two arrays of nanomagnets were patterned on the same substrate and underwent the same demagnetization process. (C) Calculated excitation energies of three types of charge ordering. The interaction energy is stronger for shorter island/spacing. The typical charge configurations for the respective Type-I, Type-II and Type-III states are shown next to the data points. (D) Population of the three types of charge configurations extracted from (A) and (B). The predicted populations for completely free (non-interactive) islands are shown for comparison. (E) Normalized populations of the three charge configurations to the predicted population for non-interactive islands. A value above 1.0 corresponds to a favorable state while unfavorable states have values below 1. The favorability clearly shows an inverse relation with the calculated excitation energies given in (C). The charge neutral Type-I configurations are strongly favored in both arrays. The Type-I ground state becomes even more favored in the array with closer charge spacing, consistent with the calculated stronger interaction energies of closer spaced charges shown in (C). These results indicate that magnetic charge ice plays a significant role in our artificial spin system.

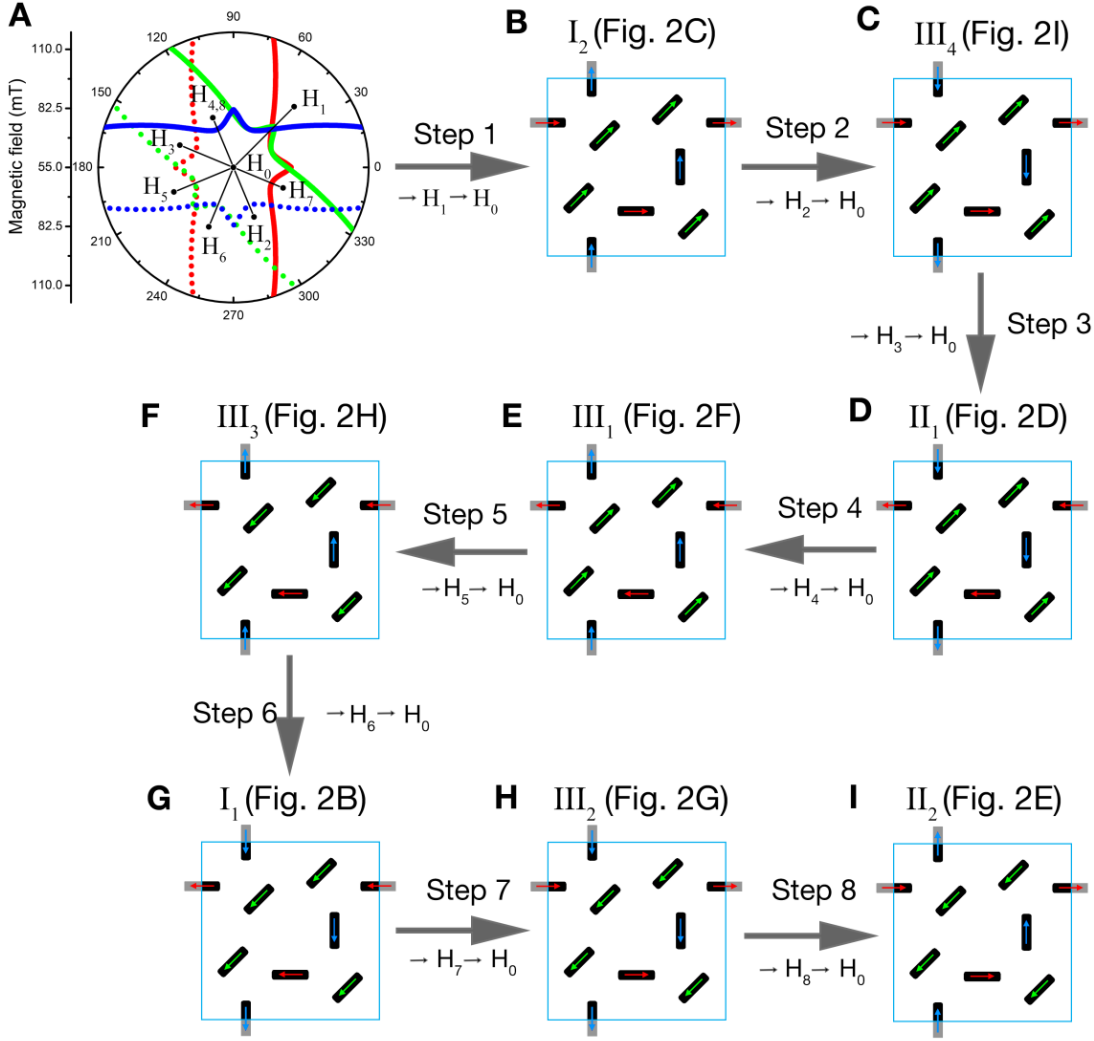

**Fig.S7. Typical protocol steps to realize all the magnetic charge ordered states.** (A) Flipping field curves for vertical (blue), horizontal (red) and diagonal (green) islands. $H_0$ – $H_8$ indicates the applied field angles and amplitudes for the eight steps shown in (B-I). (B-I) Eight steps to obtain the eight ordered sates corresponding to Figs. 2,B-I in the main text: (B) step 1: obtain order $I_2$ by applying magnetic field to $H_1$ and then reducing field to $H_0$ (zero field), (C) step 2: convert Type-$I_2$ to Type-$III_4$ by applying magnetic field to $H_2$ (flip vertical island's moment to 'down') and then reduce field to $H_0$ (zero field), (D) step 3: convert Type-$III_4$ to Type-$II_1$ by applying magnetic field to $H_3$ (flip horizontal island's moment to 'left') and then reduce field to $H_0$ (zero field), (E) step 4: convert $II_1$ to $III_1$ by applying magnetic field to $H_4$ (flip vertical island's moment to 'up') and then reduce field to $H_0$ (zero field), (F) step 5: convert $III_1$ to $III_3$ by applying magnetic field to $H_5$ (flip diagonal island's moment to 'left-down') and then reduce field to $H_0$ (zero field) (G) step 6: convert $III_3$ to $I_1$ by applying magnetic field to $H_6$ (flip vertical island's moment to 'down') and then reduce field to $H_0$ (zero field), (H) step 7: convert $I_1$ to $III_2$ by applying magnetic field to $H_7$ (flip horizontal island's moment to 'right') and then reduce field to $H_0$ (zero field), (I) step 8: convert $III_2$ to $II_2$ by applying magnetic field to $H_8$ (flip vertical island's moment to 'up') and then set field to $H_0$ (zero field).

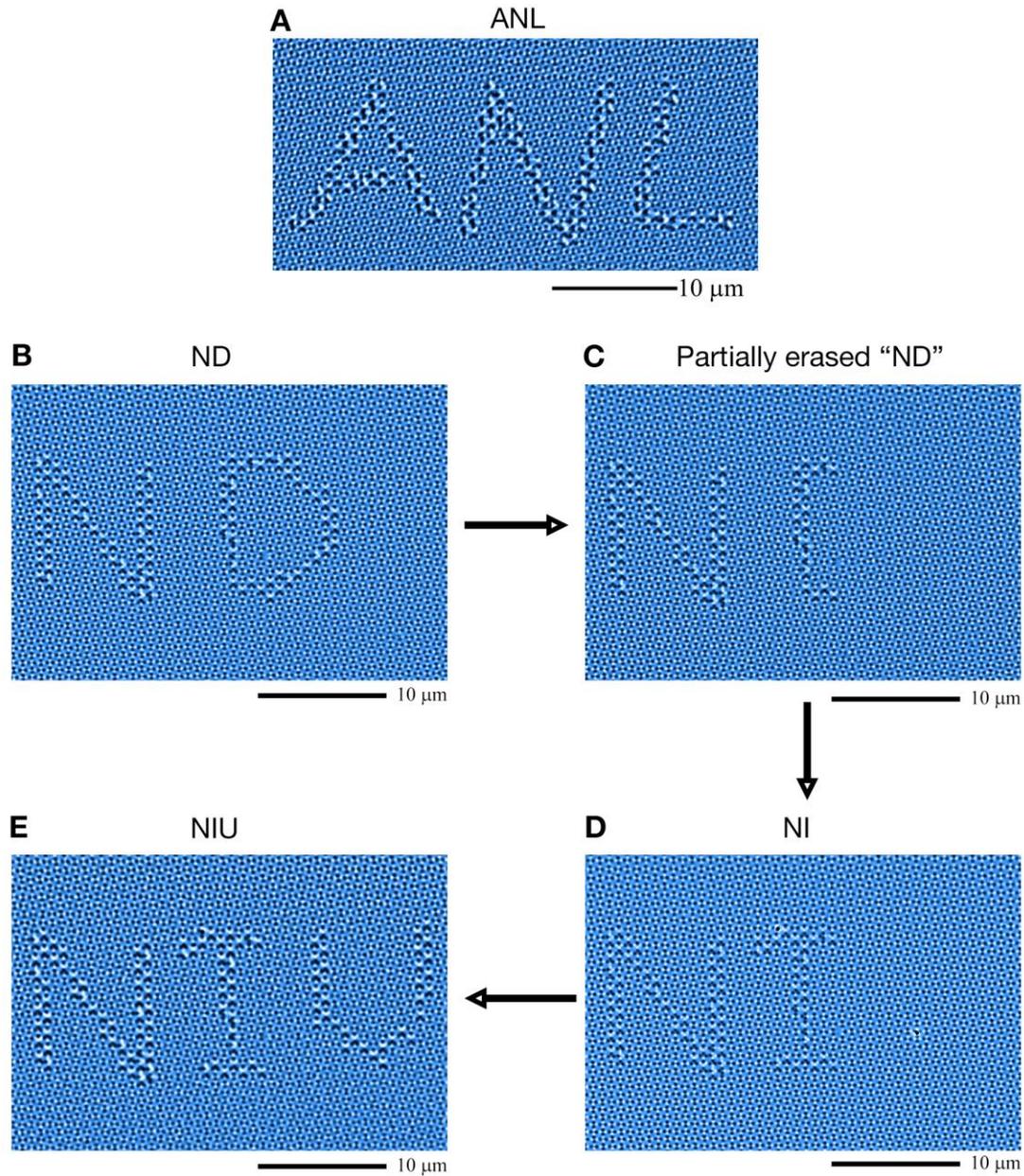

**Fig.S8. Additional MFM images demonstrating the write/read/erase functions.** (A) 'ANL' letters of Type-III states were scribed onto a Type-I background state. (B-E) Letters patterns on the same location: (B) writing letters "ND", (C) partial erasing of letter "D", (D) adding section to partially erased "D" to form letter "I", (E) adding letter "U" to form "NIU". The letters are composed of Type-III states surrounded by a background of Type-I ground states.

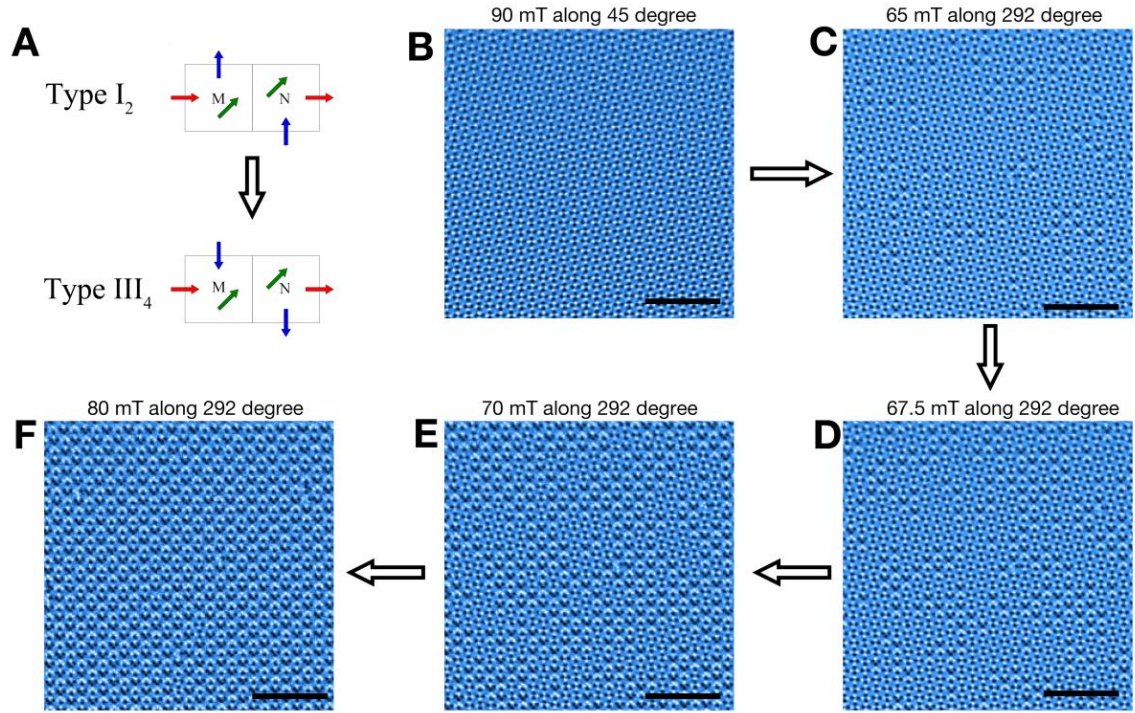

**Fig.S9. Evolution of the magnetic charge phase transition from Type-I ordering to Type-III ordering.** (A) Spin configurations of the transition from Type-$I_2$ order to Type-$III_4$ order. (B-F) MFM images delineating the gradual transition from Type-$I_2$ order (B) to Type-$III_4$ order (F) by applying magnetic field along 292 degree (see Fig. 2c) with increasing magnetic field amplitudes of 65 mT (C), 67.5 mT (D), 70 mT (E) and 80 mT (F). The black scale bar in (B-F) is 5 μm.

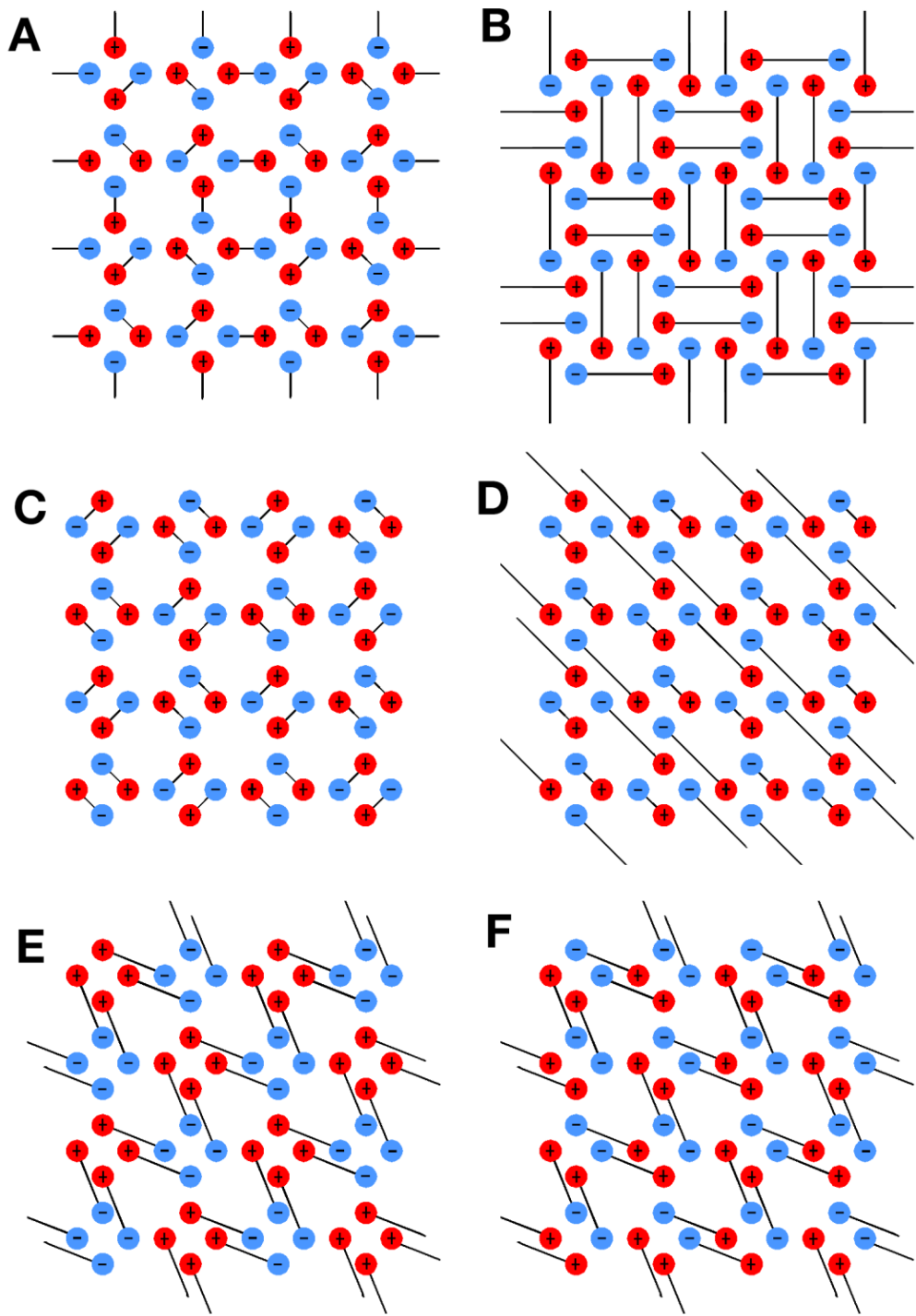

**Fig.S10. Additional designs of island arrangements for magnetic-charge patterns.**
(A-D) Various island arrangements maintaining the same magnetic charge distribution as that in the ground state of the spin structure presented in Fig. 1E and the square spin ice. (E, F) Proposed arrangement which can form type-IV (E) and type-II (F) magnetic charge order. Black lines connecting the positive and negative magnetic charges denote the island configurations.